\documentstyle[pra,aps]{revtex}

\begin{document}
\draft

\twocolumn[\hsize\textwidth\columnwidth\hsize\csname 
@twocolumnfalse\endcsname

\title{Physical interpretation of stochastic Schr\"odinger equations
in cavity QED}

\author{Tarso B.L. Kist}

\address{Instituto de F\'{\i}sica e Departamento de Biof\'{\i}sica,
Universidade Federal do Rio Grande do Sul, Cx.P. 15093,\\
91501-970 Porto Alegre RS, Brazil, and Faculty of Science,
University of Ottawa, 150 Louis-Pasteur, Ottawa, ON, Canada.}

\author{M. Orszag}

\address{Facultad de F\'{\i}sica, Pontif\'{\i}cia Universidad Cat\'olica
de Chile, Casilla 306, Santiago, Chile}

\author{T.A. Brun}

\address{Institute for Theoretical Physics, University of California, 
Santa Barbara, CA 93106-4030}

\author{L. Davidovich}

\address{Instituto de F\'{\i}sica, Universidade Federal do Rio de Janeiro,
Cx.P. 68528, 21945-970 Rio de Janeiro RJ, Brazil}

\date{\today}
\maketitle

\begin{abstract}
We propose physical interpretations for stochastic methods which have been 
developed recently to describe the evolution of a quantum system interacting 
with a reservoir. As opposed to the usual reduced density operator approach, 
which refers to ensemble averages, these methods deal with the dynamics of 
single realizations, and involve the solution of stochastic Schr\"odinger 
equations. These procedures have been shown to be completely equivalent to 
the master equation approach when ensemble averages are taken over many 
realizations. We show that these techniques are not only convenient 
mathematical tools for dissipative systems, but may actually correspond 
to concrete physical processes, for any temperature of the reservoir.  
We consider a mode of the electromagnetic field in a cavity 
interacting with a beam of two- or three-level atoms,  the field mode 
playing the role of a small system and the atomic beam standing for a 
reservoir at finite temperature, the interaction between them being given 
by the Jaynes-Cummings model.  We show that the evolution of the field 
states, under continuous monitoring of the state of the atoms which leave 
the cavity, can be described in terms of either the Monte Carlo Wave-Function
(quantum  jump) method or a stochastic Schr\"odinger equation,
depending on the system configuration. We also show that the Monte Carlo
Wave-Function approach leads, for finite temperatures, 
to localization into jumping Fock states, while the diffusion equation 
method leads to localization into states with a diffusing average photon 
number, which for sufficiently small temperatures are close approximations 
to mildly squeezed states. 
\end{abstract}

\vskip2pc] \label{sec:level1}

\section{Introduction}

The dynamics of dissipative quantum systems is frequently described through 
a master equation for the reduced density matrix, obtained by tracing out 
the degrees of freedom of the reservoir and making the Born-Markov 
approximation \cite{Cohen}. As usual in quantum mechanics, the corresponding 
solutions refer to ensembles of identical systems. In recent years, the 
attainment of low temperatures and low-dissipation regimes, as well as the 
improvement of detection techniques, has allowed the investigation of the 
dynamics of continuously monitored single quantum systems. Remarkable 
examples of these are single ions \cite{Neuhauser} or Bose-Einstein 
condensates \cite{Ketterle} in electromagnetic traps, probed by laser 
beams,  and electromagnetic fields in high-$Q$ cavities, probed by beams 
of highly excited atoms (Rydberg atoms) \cite{Brune}.  This new generation 
of experiments, combined with the difficulties usually encountered in 
solving the master equation, have stimulated the development of new 
techniques, which seek to describe the dynamics of quantum dissipative 
systems by stochastic evolutions of the corresponding state vectors 
\cite{Dalibard1,Pearle,Gisin1,Dalibard,Carmichael,Gisin,Wiseman,Wiseman2,Molmer1,Molmer2,Goetsch}. 

Basically two approaches have been proposed, one which involves random 
finite discontinuities in the dynamics of the system (Monte Carlo
Wave-Function or ``quantum jumps" method) \cite{Dalibard1,Carmichael}
and another for which the stochastic evolution is generated by a
diffusive term in the 
Schr\"odinger equation for the state vector \cite{Gisin1,Gisin}. These 
alternative techniques can be shown to be formally equivalent to the  
master equation approach \cite{Dalibard1,Wiseman}, and in some cases do 
lead to a dynamical behavior resembling the experimental monitoring of 
a single realization. However, they are often regarded as mathematical 
tricks, with no relation to a concrete physical evolution of the system. 
The advantage of using them, from the numerical point of view,  is that one 
deals with state vectors, instead of density matrices, thus reducing 
the total amount of matrix elements to be calculated. In addition, they may 
provide insights into the behavior of dissipative systems. In fact, and 
because of these two points, they have been extensively applied to 
dissipative quantum systems, especially in the fields of quantum optics 
\cite{Dalibard,Gisin,Wiseman,Wiseman2,Molmer1,Molmer2,Garraway,Adriano,Tarso,Zelaquett,Cresser} 
and solid state physics \cite{Imamoglu}. In some cases, these methods have 
led to analytical descriptions of the dissipation process 
\cite{Garraway,Tarso,Burt}. More recently, the Monte Carlo Wave-Function 
(MCWF) method has been extended to non-Markovian interactions and to strong 
reservoir couplings beyond the Born and rotating-wave approximations 
\cite{Imamoglu}, and also to nonlinear master equations 
\cite{Molmer1,Molmer2}. 

In this work, we show that it is  possible to interpret the above-mentioned
stochastic evolutions in terms of continuous measurements made on concrete 
physical systems, for any temperature of the reservoir. This paper extends 
therefore the pioneer work of Wiseman and Milburn \cite{Wiseman,Wiseman2}, 
who developed physical interpretations for Schr\"odinger stochastic 
equations describing the evolution of a cavity mode in contact with a 
zero-temperature reservoir. They showed that the mode of the 
electromagnetic field  is described by a quantum jump equation if the 
outgoing light is directly detected by a photodetector, while homodyne 
or heterodyne detection schemes lead to two different stochastic 
Schr\"odinger equations for the state of the field.  

We also consider 
the time-dependent behavior of  a mode of the electromagnetic field in a 
cavity interacting with a reservoir (which may be associated to the 
continuum of modes of the field outside of the cavity, to which the 
internal field may be coupled via a semi-transparent mirror). Our 
interpretations rely on the fact that the same master equation is 
obtained for quite different reservoir models. We take thus as the 
reservoir an atomic beam which crosses the cavity, interacting resonantly 
with the field inside it. We show that the Monte Carlo or quantum jump 
method can be reproduced by taking as reservoir a beam of continuously 
monitored two-level atoms, prepared initially in a mixture of the two 
resonant states, in such a way that the population ratio between the 
excited and the ground state equals the Boltzman factor (see Fig.\ \ref{fig1}).
On the other hand, a Schr\"odinger diffusion equation is obtained when the 
reservoir is assumed to be made of three-level atoms with a 
twofold-degenerated ground state, prepared in the same kind of 
statistical mixture as before, but now with the two ground state levels 
placed in a coherent superposition. While a resonant exchange of energy 
is allowed between the cavity mode, the excited state and one of the 
ground state levels, the transition between the other ground state level 
and the excited state is assumed to be driven by an external (essentially 
classical) field (see Fig. \ref{fig2}). 

In order to further explore the physics underlying these approaches, we 
calculate the evolution of an initial field in the cavity under continuous 
monitoring, for different types of initial states of the field, with the 
two different methods mentioned above. We find that for temperatures 
different from zero the two approaches lead to two kinds of localization 
in state space. For the quantum jump method, the state of the system 
approaches a Fock state, which suffers quantum jumps in such a way that 
the average distribution in time of the number of photons satisfies the 
thermal distribution. On the other hand, the diffusion equation leads to 
states which, for sufficiently low temperatures (average number of thermal 
photons smaller than one), are quite close to mildly squeezed states. Even 
though localized in phase space, these states have a diffusive behavior, 
spanning eventually a region of the phase space in such a way that again 
the time-averaged photon statistics coincides with the thermal distribution.
We will show that our atomic model for the reservoir allows a simple 
interpretation of these localization phenomena, which extends to finite 
temperatures the discussion made by Garraway and Knight \cite{Garraway}, 
and illustrates the  
general localization properties of quantum state diffusion equations  
demonstrated by Gisin and Percival \cite{Gisin}.

In Section II, we review the stochastic approach to dissipative systems.
In Section III we propose a physical interpretation for the Monte Carlo 
quantum jump approach to the problem of field dissipation in cavity QED, 
for any reservoir temperature, while in Section IV we show how to interpret
physically a description of the same problem based on a stochastic 
Schr\"odinger equation. In Section V, we display our numerical results, 
and show that, depending on the physical procedure used to monitor 
continuously the field in the cavity, one may get localization in state 
space. Our conclusions are summarized in Section VI, while details of the 
calculations are displayed in the Appendices.

\section{Stochastic Schr\"odinger equations and dissipative systems}

A wide class of master equations describing the evolution of dissipative 
quantum systems can be written in the Lindblad form \cite{Lindblad}: 
\begin{equation}\label{master1}
\dot{\rho}_{\rm S} ={\cal L}\rho_{\rm S}\,,
\end{equation}
where
\begin{equation}
{\cal L}={\cal L}_0+\sum_n {\cal L}_n\,,
\end{equation}
\begin{equation}
{\cal L}_0\rho_{S}= \frac{i}{\hbar}[\rho_{\rm S},H_{\rm s}]\,,
\end{equation}
\begin{equation}
{\cal L}_n\rho_{\rm S}=-\frac{1}{2}[C_n^{\dagger}C_n\rho_{\rm S} + 
\rho_{\rm S}
C_n^{\dagger}C_n] + C_n\rho_{\rm S}C_n^{\dagger}\,,
\end{equation}
$\rho_{\rm S}$ is the reduced density operator for the ``small'' system 
$S$ (obtained by tracing out the degrees of freedom of the reservoir $R$ 
from the density  operator for the full system $S+R$), and $H_{\rm S}$ 
describes the Hamiltonian evolution of the small system $S$ in the 
interaction picture. The operators $C_n$ act on the space of states of 
the small system  $S$, and express the interaction of $S$ with the 
reservoir $R$. The number of them depends on the nature of the problem. 

An example of such an equation is the master equation for a field in a 
lossy cavity, at temperature $T$, given in the interaction picture by
\begin{eqnarray}  \label{masterfield}
\frac{d\rho _f}{dt}&=&\Gamma \overline{n}(a^{\dagger }\rho _fa-\frac 12%
aa^{\dagger }\rho _f-\frac 12\rho _faa^{\dagger })\nonumber \\ 
&+&\Gamma (1+\overline{n})(a\rho _fa^{\dagger }-\frac 12a^{\dagger }a^{}\rho
_f-\frac 12\rho _fa^{\dagger }a)\,,
\end{eqnarray}
where $a$ and $a^\dagger$ are the photon annihilation and creation 
operators, respectively, $\overline{n}$ is the average number of thermal 
photons, given by Planck's distribution, and $\Gamma=1/t_{\rm cav}$, 
where $t_{\rm cav}$ is the damping time.
In this case, one could set 
\begin{equation}
C_1\equiv\sqrt{\Gamma(1+\overline{n})}a,\qquad C_2\equiv\sqrt{\Gamma
\overline{n}}a^\dagger\,.
\end{equation}

A formal solution of Eq. (\ref{master1}) is
\begin{equation}\label{formal1}
\rho(t)=\exp\left({\cal L}t\right)\rho(0)\,.
\end{equation}
Let us define
\begin{equation}\label{Jn}
J_n\rho=C_n\rho C_n^\dagger\,,
\end{equation}
and write 
\begin{equation}\label{formal2}
\rho(t)=\exp\left\{{\cal L}_0t+\sum_n\left[J_n+\left({\cal L}_n-J_n\right)
\right]t\right\}\rho(0)\,.
\end{equation}
Note that
\begin{equation}\label{L-J}
\left({\cal L}_n-J_n\right)\rho_{\rm S}=-\frac{1}{2}\left(C_n^{\dagger}
C_n\rho_{\rm S} + \rho_{\rm S}C_n^{\dagger}C_n\right)\,.
\end{equation}

Applying Dyson's expansion to Eq.\ (\ref{formal2}), we get:
\begin{eqnarray}\label{Dyson1}
\rho(t)&=&\sum_{m=0}^\infty\int_0^t dt_m\int_0^{t_m}dt_{m-1}\dots
\int_0^{t_2}dt_1\nonumber\\
&&\{S(t-t_m)(\sum_nJ_n)S(t_m-t_{m-1})\nonumber\\
&&\times\dots(\sum_nJ_n)S(t_1)\}\rho(0)\,,
\end{eqnarray}
where
\begin{equation}\label{S}
S(t)=\exp\left\{\left[{\cal L}_0+\sum_n\left({\cal L}_n-J_n\right)
\right]t\right\}\,.
\end{equation}
Equation (\ref{Dyson1}) may be rewritten in the following way:
\begin{eqnarray}\label{trajectories}
\rho(t)&=&\sum_{m=0}^\infty\sum_{\{n_i\}}\int_0^t dt_m\int_0^{t_m}
dt_{m-1}\dots\int_0^{t_2} dt_1\nonumber\\
&&\{S(t-t_m)J_{n_m}S(t_m-t_{m-1})\nonumber\\
&&\times\dots J_{n_1}S(t_1)\}\rho(0)\,.
\end{eqnarray}

Each term in the above double sum can be considered as a quantum 
trajectory, the reduced density operator at time $t$ being given by the 
sum over all possible quantum trajectories \cite{Carmichael}. For each 
of these trajectories, Eq. (\ref{trajectories}) shows that the evolution of 
the system can be considered as a succession of quantum jumps, associated 
to the operators $J_n$, interspersed by smooth time evolutions, associated 
with the operators $S(t)$. The probability of each trajectory is given by 
the trace of the corresponding term in Eq. (\ref{trajectories}).

From Eqs. (\ref{L-J}) and (\ref{S}), we can write:
\begin{equation}
S(t)\rho=N(t)\rho N(t)^\dagger\,,
\end{equation}
where
\begin{equation}\label{N}
N(t)=\exp\left[-{i\over\hbar}H_St-{t\over2}\sum_n\left(C_n^\dagger 
C_n\right)\right]\,.
\end{equation}
Therefore, if $\rho$ is a pure state, then $S(t)\rho$ is also a pure 
state. The same is true for  $J_n\rho$, with $J_n$ defined by Eq. (\ref{Jn}). 
This implies that a pure state remains pure, when a single quantum 
trajectory is considered. Note also that the evolution between jumps is 
given by the non-unitary operator $N(t)$.

It is clear from Eq. (\ref{formal2}) that different choices of the jump 
operators are possible. These different choices  correspond to different 
decompositions in terms of quantum trajectories of the time evolution of 
the density operator $\rho_S$ and, eventually, to different experimental 
schemes leading to the continuous monitoring of the evolution of the 
system. It is precisely due to this continuous monitoring that an initial 
pure state remains pure, since no information is lost in this situation: 
for a field in a cavity, this continuous monitoring amounts to accounting 
for every photon gained or lost by the field, due to its interaction with 
the reservoir. 

We will discuss now two different realizations of the reservoir, for a 
field in a cavity, which will lead to a Monte Carlo quantum jump equation, 
for the first realization, and to a Schr\"odinger equation
with stochastic terms, for the second one.

\section{Simulation of a Monte Carlo SSE.}

We exhibit in this Section a physical realization of the Monte Carlo 
method. The corresponding experimental scheme is shown in Fig.\ \ref{fig1}. 
A monokinetic atomic beam plays the role of
a reservoir $R$ and crosses a lossless cavity, interacting with one mode
of the electromagnetic field. The cavity mode plays the role
of a small system $S$.
The atoms, regularly spaced along the atomic beam, are prepared in one
of  two Rydberg states: an upper state $|a\rangle $ or a lower state 
$|b\rangle $. The transition frequency $\omega$ between these
two states is assumed to be resonant with the cavity mode. A similar 
model of reservoir was adopted in Section 16.1 of Ref.\ \cite{Scully}.

The state of the atoms is measured by a detector just at the
exit of the cavity. The ratio between the flux of upper state atoms $r_{a}$
and the lower state atoms $r_{b}$ before their entrance into the cavity 
is chosen so that 
\begin{equation}  \label{distribT}
\frac{r_{a}}{r_{b}}=e^{-\hbar \omega/k_{B}T}\equiv \frac{\overline{n}}{1+%
\overline{n}},
\end{equation}
where $\hbar \omega$ is the difference in energy between $|a\rangle $ and 
$%
|b\rangle $, and, as will be shown in the next paragraphs, $T$ is the 
reservoir temperature.  The constant $k_{B}$ represents the Boltzmann 
constant and $%
\overline{n}$, given by Planck's formula ($\overline{n}=[exp(\hbar
w/k_{B}T)-1]^{-1}$), is the mean occupation of the modes with energy $%
\hbar w$ in a bath at temperature $T$. 

We analyze now the time evolution of the state
vector $|\Psi (t)\rangle $ of  $S$, under the continuous measurement of 
the atoms after they leave the cavity.  We also assume that one knows the 
state of each atom before it interacts with the cavity. This may be 
achieved by selectively exciting the atoms to $|a\rangle$ or $|b\rangle$, 
according to the proportion given by Eq. (\ref{distribT}).  We will adopt the
following simplifying assumptions: (a) the atom-field interaction time 
$\tau $ is the same for all atoms; (b) the spatial profile of the electric 
field is constant; (c) the cavity is perfect, $i.e.$, the field state is 
changed only by the atoms; (d) the atom-field coupling constant $\lambda $ 
and the interaction time $\tau 
$ are both small, so that the atomic state rotation is very small; (e) the
rotating-wave and dipole approximations will be used; and (f), according to
the statements (d) and (e), quantum cooperative effects will be
neglected. In this case the interaction Hamiltonian in the interaction
picture will be: 
\begin{equation}  \label{schwer}
H=\hbar \lambda \left( |b\rangle \langle a|a^{\dagger }+|a\rangle \langle
b|a\right) .
\end{equation}
The operators $a$ and $a^{\dagger }$ are annihilation and creation 
operators, acting on the space of states of the field
mode. Just before the $i$th atom enters the cavity, the state describing 
the combined system (atom $i$ $+$ field) is given by
\begin{equation}
|\Psi _{a-f}(t_{i})\rangle =|\Psi (t_{i})\rangle \otimes
|\Psi_{a}(t_{i})\rangle .
\end{equation}
Here $|\Psi _{a}(t_{i})\rangle =|a\rangle $ or $|\Psi _{a}(t_{i})\rangle
=|b\rangle $, depending on the state to which the atom was excited just 
prior to entering the cavity.  

{\it {\rm At time $t_{i}+\tau $, the atom-field state vector, up to second
order in $\tau $, is given by: 
\begin{eqnarray}\label{evolT}
&&|\tilde{\Psi}_{a-f}(t_{i}+\tau )\rangle =\left( 1-i\lambda \tau |b\rangle
\langle a|\,a^{\dagger }-i\lambda \tau |a\rangle \langle b|\,a\frac {} {}%
\right.  \nonumber  \\
&&-\left. \frac{\lambda ^{2}\tau ^{2}}{2}|b\rangle \langle b|\,a^{\dagger }a-%
\frac{\lambda ^{2}\tau ^{2}}{2}|a\rangle \langle a|\,aa^{\dagger }\right)
|\Psi _{a-f}(t_{i})\rangle ,
\end{eqnarray}
where the tilde indicates that the state vector is not normalized. The
expansion (\ref{evolT}) should be very good in view of condition (d). We 
assume that $(r_a+r_b)\tau<1$,  so that there is at most one atom inside 
the cavity at each instant of time. After this atom exits the cavity and 
is detected, one of the following four cases will be realized:\\
$i$. The atom enters the cavity in state $|b\rangle $ and  is
detected in the same state. In this case, according to Eq. (\ref
{evolT}), the state of $S$ at time $t=t_{i}+\tau $ will be given by
\begin{equation}  \label{caso1}
|\tilde{\Psi}(t_{i}+\tau )\rangle =\left( 1-\frac{\lambda ^{2}\tau ^{2}}{2}%
a^{\dagger }a\,\right) |\Psi (t_{i})\rangle .
\end{equation}
$ii$. The atom enters the cavity in state $|a\rangle $ and it is detected in
the same state $|a\rangle $. In this case, 
\begin{equation}  \label{caso2}
|\tilde{\Psi}(t_{i}+\tau )\rangle =\left( 1-\frac{\lambda ^{2}\tau ^{2}}{2}%
aa^{\dagger }\,\right) |\Psi (t_{i})\rangle .
\end{equation}
$iii$. The atom enters the cavity in the state $|b\rangle $ and it is
detected in the state $|a\rangle $. In this case, 
\begin{equation}  \label{caso3}
|\tilde{\Psi}(t_{i}+\tau )\rangle =-i\lambda \tau a\,|\Psi
(t_{i})\rangle .
\end{equation}
$iv$. The atom enters the cavity in the state $|a\rangle $ and it is
detected in the state $|b\rangle $. Then,
\begin{equation}  \label{caso4}
|\tilde{\Psi}(t_{i}+\tau )\rangle =-i\lambda \tau a^{\dagger }\,|\Psi
(t_{i})\rangle .
\end{equation}
}}

{\it {\rm Note that in the cases $i$ and $ii$ a small change in the state of
``S'' takes place, while in the cases $iii$ and $iv$ a big change may happen 
(quantum jump). However, these last two cases are very rare, due to the small
change of the atomic state during the interaction time. 

We consider now the change of  $|\Psi \rangle $ from $t$ to $t+\delta t$, 
where the time interval $\delta t$ is large enough so that many atoms go 
through the cavity during this time interval ($n_a=r_a\delta t\gg1$, 
$n_b=r_b\delta t\gg1$), and also much smaller than $t_{\rm cav}/
\overline{n}\langle n\rangle$, where $\langle n\rangle$ is the average 
number of photons in the state. This last condition, as it will be seen 
later, implies that the probability of a quantum jump during $\delta t$ is 
very small. In most of the time intervals $%
\delta t$ the atoms will be detected at the same state they came in, since 
the transition probability is very small. The evolution of $%
|\Psi \rangle $ during these intervals will be given by: 
\begin{eqnarray}\label{nojump}
|\tilde{\Psi}(t+\delta t)\rangle &=&\left( 1-\frac{\lambda ^{2}\tau ^{2}}{2}%
aa^{\dagger }\right) ^{n_{a}}\nonumber\\
&\times&\left( 1-\frac{\lambda ^{2}\tau ^{2}}{2}%
a^{\dagger }a\right) ^{n_{b}}|\Psi (t)\rangle  \nonumber  \\
&=&\left( 1-\frac{n_{a}\lambda ^{2}\tau ^{2}}{2}aa^{\dagger }-\frac{%
n_{b}\lambda ^{2}\tau ^{2}}{2}a^{\dagger }a\right) |\Psi (t)\rangle .
\end{eqnarray}
This result does not depend on the ordering of the upper-state and 
lower-state atoms.  We also note that in the interaction picture the 
state vector does not evolve when there is no atom inside the cavity, 
since the only source of field dissipation is the interaction with the 
atomic beam. 

Equation (\ref{nojump}) displays the interesting property that the 
wave-function of the system (and, consequently, the mean energy) may 
change even when there is no exchange of energy between the system and 
the measurement apparatus (represented by the atoms in the present case).
An easy way to understand this effect physically is to imagine that all 
atoms are sent into the cavity in the lower state, and are detected in 
the same state after exiting the cavity, for a given realization of the 
system, which starts with a coherent state in the cavity. Then, even 
though there is no exchange of energy between the atoms and the field 
in the cavity, as time evolves the ground state component of the initial 
state should also increase, since the results of the measurements lead 
to an increasing probability that there is a vacuum state in the cavity. 
In other words, the fact that there is no quantum jump, for that specific 
trajectory, provides us with information about the quantum state of the 
system, and this information leads to an evolution of the state. This is 
closely related to the quantum
theory of continuous measurement \cite{Davies,Ueda} and also to quantum
non-demolition measurement schemes proposed recently \cite{L}. This problem
is also very similar to that of a Heisenberg microscope in which even the
unsuccessful events of light scattering produce a change in the
quantum-mechanical state of the particle \cite{Dicke}. 

We introduce now the following definitions: 
\begin{equation}  \label{vinho}
\Gamma \equiv (r_b-r_a)\lambda ^2\tau ^2=\frac{r_b}{1+\overline{n}}%
\lambda ^2\tau ^2=\frac{r_a}{\overline{n}}\lambda ^2\tau ^2,
\end{equation}
\begin{equation}  \label{C1C2}
C_1\equiv \sqrt{\Gamma (1+\overline{n})}\,a,\;\;\;\;\;\;C_2\equiv \sqrt{%
\Gamma \overline{n}}\,a^{\dagger }\,.
\end{equation}
Using these definitions and relation (\ref{distribT}),  Eq. (\ref{nojump}) 
may be rewritten in the following way;
\begin{equation}  \label{evtem}
|\tilde \Psi (t+\delta t)\rangle =\left[ 1-\frac{\delta t}{2}
\sum_mC_m^{\dagger
}C_m\right] |\Psi (t)\rangle .
\end{equation}

If an atom enters the cavity in state $|a\rangle$ and is detected in the 
state $|b\rangle $, the state vector of $S$ suffers a ``quantum jump'', 
and one photon is added to that system. On the other hand, a 
deexcitation in $S$ occurs if an atom which entered in $|b\rangle $ is
detected in the state $|a\rangle $. The probability of this events to occur
may be calculated by using Eqs. (\ref{C1C2}) and (\ref{caso3}) or (\ref
{caso4}); thus, the probability of an excitation (action of $a^{\dagger }$)
to occur between $t$ and $t+\delta t$ is given by: 
\begin{equation}
\delta p_1=\delta t\langle \Psi (t)|C_1^{\dagger }C_1|\Psi (t)\rangle .
\end{equation}
The probability of a deexcitation (action of $a$) during this time
interval is: 
\begin{equation}
\delta p_2=\delta t\langle \Psi (t)|C_2^{\dagger }C_2|\Psi (t)\rangle .
\end{equation}
The probabilities $\delta p_1$ and $\delta p_2$ are very low, so that the
joint probability of having one excitation and one deexcitation during the 
same
time interval $\delta t$ is negligible. One may therefore write:
\begin{equation}  \label{quase}
|\tilde \Psi (t+\delta t)\rangle =C_1^{\delta N_1}C_2^{\delta N_2}
\left[ 1-\frac{\delta t}2%
\sum_mC_m^{\dagger }C_m\right] |\Psi (t)\rangle\,.
\end{equation}
where $\delta N_1$ and $\delta N_2$ are equal to one or zero, with 
probabilities $%
\delta p_1$ and $\delta p_2$ for  $\delta N_1$ and $\delta N_2$ to be
equal to one, respectively. This may be represented by writing the 
statistical mean $M(\delta N_m)=\langle C_m^\dagger C_m\rangle\delta t$. 
Also, $\delta N_m\delta N_n=\delta N_m\delta_{nm}$. One should note that 
the instants of time in which the quantum jumps occur during the time 
interval $\delta t$ are irrelevant, since the jump operators can be 
commuted through the no-jump evolution, the commutation producing an 
overall phase which goes away upon renormalization of the state. This 
can be easily seen by rewriting the no-jump evolution, during a time 
interval $\delta t_j<\delta t$,  as an exponential:
\begin{eqnarray}
1-\frac{\delta t_j}2\sum_mC_m^{\dagger }C_m&=& \exp\left(-\frac{\delta 
t_j}2\sum_mC_m^{\dagger }C_m\right)\nonumber\\
&+&O[(\delta t_j)^2]\,,
\end{eqnarray}
and using that 
\begin{equation}
C_i e^{-\frac{\delta t_j}2\sum_mC_m^{\dagger }C_m}= e^{-\frac{\delta t_j}2
\sum_mC_m^{\dagger }C_m}C_i e^{\lambda_i}\,,
\end{equation}
where $\lambda_1=-(\delta_j/2)\Gamma(1+\overline{n})$ and $\lambda_2=
(\delta t_j/2)\Gamma \overline{n}$. 

The results of the measurement may be simulated by picking 
random numbers. The state vector given by Eq. (\ref{quase}) may be
normalized as follows:
\begin{eqnarray}\label{quase2}
\mid \psi (t+\delta t)\rangle &=&\left[\frac{C_{1}}{\sqrt{C_{1}^{\dagger }
C_{1}}}%
\delta N_{1}+\frac{C_{2}}{\sqrt{C_{2}^{\dagger }C_{2}}}\delta N_{2}\right. 
\nonumber\\
&+&(1-\delta N_{1})(1-\delta N_{2})\left(1-\frac{\delta t}{2}%
\sum_{m}C_{m}^{\dagger }C_{m}\right)\nonumber\\
&\times& \left. \left( 1-\delta t\sum_{m}\langle
C_{m}^{\dagger}C_{m}\rangle \right) ^{-\frac{1}{2}} \right] \mid \psi (t)\rangle. 
\end{eqnarray}

In the above equation, the first two terms represent the possible jumps,
each normalized, as in the Monte Carlo method, and the last term is the
no-jump evolution contribution, normalized with the corresponding
prefactor that rules out the jumps. From Eq. (\ref{quase2}) one gets for 
$\mid d\psi (t)\rangle \equiv \mid \psi (t+\delta t)\rangle -\mid \psi
(t)\rangle$: 
\begin{eqnarray}\label{dpsi}
\mid d\psi (t)\rangle &=&\left\{\sum_{m}\left[ \frac{C_{m}}
{\sqrt{C_{m}^{\dagger
}C_{m}}}-1\right] \delta N_{m}\right.\nonumber \\
&-&\frac{\delta t}{2}\sum_{m}(C_{m}^{\dagger }C_{m}-\langle C_{m}^{\dagger
}C_{m}\rangle )\Bigg\}\mid \psi (t)\rangle . 
\end{eqnarray}

\section{Simulation of the homodyne SSDE.}

We show now that, by a suitable modification of the atomic configuration, 
it is also possible to interpret physically the stochastic Schr\"odinger 
equations in terms of continuous measurements made on atoms which cross 
the cavity containing the field. The corresponding scheme is shown in 
Fig.\ 2: a beam of three-level atoms with a degenerate lower state 
(states $b$ and $c$) crosses the cavity, the field in the cavity being 
resonant with a transition between one of the two lower levels (say, 
level $b$) and the upper atomic state $a$, while a strong classical 
field connects the other lower state with the upper level (one may 
assume that both fields are circularly polarized, so that the cavity 
field cannot connect $a$ and $c$, while the strong field does not induce
 transitions between $a$ and $b$). 

We also assume that the atom is prepared in either a coherent superposition 
of the two lower levels:
\begin{equation}  \label{xi1}
\mid \psi _{atom}\rangle =\frac 1{\sqrt{2}}(\mid b\rangle +\mid c\rangle ),
\end{equation}
or in the upper one, following a Boltzmann distribution corresponding to a
temperature T for the atoms, which act as a reservoir for the quantum field 
in the cavity.

In the interaction picture, one can write: 
\begin{eqnarray}  \label{xi2}
H&=&\hbar g_{ac}(\varepsilon \mid a\rangle \langle c\mid +\varepsilon
\mid c\rangle \langle a\mid )\nonumber \\ 
&+&\hbar g_{ab}(a^{\dagger }\mid b\rangle \langle a\mid +a\mid a\rangle
\langle b\mid .
\end{eqnarray}
We assume for simplicity that $g_{ac}=g_{ab}=g$, and that $\varepsilon$ is 
real. The time evolution of the wave function , to second order in the 
coupling constant is: 
\begin{equation}  \label{xi3}
\mid \psi _{}(t+\tau )\rangle =\left[ 1-\frac{iH\tau }\hbar 
-\frac{H^2\tau ^2%
}{2\hbar ^2}\right] \mid \psi _{}(t)\rangle .
\end{equation}

As in the previous model, there are two possible quantum jump processes. 
The first one corresponds to the atom entering the cavity in the coherent 
superposition of  lower states, and being detected in the upper state. 
After the measurement, the state of the field is given by:
\begin{equation}  \label{xi4}
\mid \psi _{}(t+\tau )\rangle _f^{(b,c\rightarrow a)}=\frac{-ig\tau }
{\sqrt{2%
}}(\varepsilon +a)\mid \psi _{}(t)\rangle _f\,.
\end{equation}

The corresponding probability of detecting an atom in $\mid a\rangle $, 
after a time interval $\delta t$, staring from the initial superposition 
state, is given by: 
\begin{equation}  \label{xi5}
\delta p_1=n_b\frac{g^2\tau ^2}2\langle \psi _f(t)\mid (\varepsilon
+a^{\dagger })(\varepsilon +a^{})\mid \psi _f(t)\rangle ,
\end{equation}
where $n_b\equiv r_b\delta t$, $r_b$ being the rate of atoms injected in
the superposition of the lower states.

The second jump process corresponds to the atom entering the cavity in 
the upper state $|a\rangle$, and being detected in the superposition of 
lower states. Then, the state of the field after the measurement is:
\begin{equation}  \label{xi6}
\mid \psi _{}(t+\tau )\rangle _f^{(a\rightarrow b,c)}=\frac{-ig\tau }
{\sqrt{2%
}}(\varepsilon +a^{\dagger })\mid \psi _{}(t)\rangle _f\,.
\end{equation}

The corresponding probability is given by:
\begin{equation}  \label{xi7}
\delta p_2=n_a\frac{g^2\tau ^2}2\langle \psi _f(t)\mid (\varepsilon
+a^{})(\varepsilon +a^{\dagger })\mid \psi _f(t)\rangle \,,
\end{equation}
where $n_a=r_a\delta t$ is the number of atoms which enter the cavity in 
state $|a\rangle$, during the time interval $\delta t$. 

This analysis suggests that the quantum jump operators corresponding to 
these two processes should be, respectively,
\begin{mathletters}  \label{xi8}
\begin{equation}
C_1= \sqrt{\Gamma (1+\overline{n})}(\varepsilon +a)
\end{equation}
and
\begin{equation}
C_2=\sqrt{\Gamma \overline{n}}(\varepsilon +a^{\dagger }),
\end{equation}
\end{mathletters}
where 
\begin{equation}  \label{vinho2}
\Gamma \equiv (r_b-r_a)\frac{g^2\tau ^2}{2}=\frac{r_b}{1+\overline{n}}%
\frac{g^2\tau ^2}{2}=\frac{r_a}{\overline{n}}\frac{g^2\tau ^2}{2},
\end{equation}

Formally, these jump operators are retrieved by rewriting the Master 
Equation (\ref{masterfield}) in the following equivalent form:
\begin{eqnarray}
&&\frac{d\rho _{f}}{dt}=(J_{1}+J_{2})\rho _{f}-\frac{\Gamma 
(1+\overline{n})}{2}\left[ (a^{\dagger }a+2\varepsilon
a+\varepsilon ^{2})\rho _{f}\right.\nonumber\\
&+&\left.\rho _{f}(a^{\dagger }a+2\varepsilon a^{\dagger
}+\varepsilon ^{2})\right]-\frac{\Gamma \overline{n}}{2}\left[ 
(aa^{\dagger }+2\varepsilon
a^{\dagger }+\varepsilon ^{2})\rho _{f}\right. \nonumber\\
&+&\left.\rho _{f}(aa^{\dagger }+2\varepsilon
a^{{}}+\varepsilon ^{2})\right],
\end{eqnarray}
with: 
\begin{equation}
\begin{array}{c}
J_{i}=C_{i}\rho C_{i}^{\dagger }, \\ 
i=1,2
\end{array}
\label{xi11}
\end{equation}
being associated with the jumps, the operators $C_i$ being now given by 
Eqs. (\ref{xi8}).

We derive now the stochastic Schr\"odinger equation that describes the 
present measurement scheme.

With the above jump operators, and using the expansion given by Eq.\ (\ref
{trajectories}), we show in the appendix A that the joint probability of
getting $m_{1}$ and $m_{2}$ jumps corresponding respectively to the first 
and second processes described above is given by the
following expression: 
\begin{eqnarray}\label{xi19}
P_{m_{1},m_{2}}(\Delta t) &=&\left[ \exp \mu _{1}\frac{(\mu _{1})^{m_{1}}}{%
m_{1}!}\right] \left[ \exp \mu _{2}\frac{(\mu _{2})^{m_{2}}}{m_{2}!}\right] 
 \nonumber\\
&\times&{\rm Tr}\left\{ 
\exp \beta ^{\prime }\left[1+\frac{1}{\varepsilon }(m_{1}a+m_{2}a^{\dagger
})\right]\rho\right.\nonumber  \\ 
&\times&\left.\left[1+\frac{1}{\varepsilon }(m_{1}a^{\dagger }+m_{2}a)
\right]\exp \beta ^{\dagger
\prime }
\right\} ,
\end{eqnarray}

where: 
\begin{eqnarray}
\mu _{1} &=&\Gamma \Delta t\varepsilon ^{2}(1+\overline{n}),  
\label{xi20} \\
\mu _{2} &=&\Gamma \Delta t\varepsilon ^{2}(\overline{n}),  \nonumber \\
\beta ^{\prime } &=&-\frac{\Gamma \Delta t}{2}\left[ a^{\dagger }a(2%
\overline{n}+1)+2\varepsilon a(\overline{n}+1)+2\varepsilon a^{\dagger }%
\overline{n}+\overline{n})\right] .  \nonumber
\end{eqnarray}

From Eqs.\ (\ref{xi19}) and (\ref{xi20}), one can readily find $\langle
m_{i}\rangle $ and $\langle m_{i}^{2}\rangle $ for i=1,2.

Up to order $\varepsilon^{-3/2}$, one finds: 
\begin{eqnarray}
\langle m_{i}\rangle &=&\mu _{i}\left(1+\frac{2}{3}\langle X_{1}\rangle 
\right),
\label{xi22}\nonumber \\
\langle m_{i}^{2}\rangle &=&\mu _{i}, 
\end{eqnarray}
with
\begin{equation}
X_{1} \equiv \frac{a+a^{\dagger }}{2}. 
\end{equation}

Going back to the definition of  $S(t)$, one may write: 
\begin{equation}
S(\Delta t)=N(\Delta t)\rho N^{\dagger }(\Delta t),  \label{xi23}
\end{equation}
in terms of a smooth evolution operator $N$ that preserves pure states.
This operator $N$ is given by Eq.\ (\ref{N}). with the jump operators 
$C_m$ now given by Eqs. (\ref{xi8}). 
Now, if we consider a sequence of jumps (of the two kinds, in the present
analysis) and evolutions, the state vector of the field will evolve
according to: 
\begin{eqnarray} \label{xi24}
&&\mid\widetilde{\psi }\rangle _{f}(\Delta t)=N(\Delta
t-t_{m})C_{2}N(t_{m}-t_{m-1})C_{1}...\mid \psi \rangle _{f}(0)\nonumber \\
&&= N(\Delta t)C_{2}^{m_{2}}C_{1}^{m_{1}}\mid \psi \rangle _{f}(0). 
\end{eqnarray}

In the last step, in deriving Eq.\ (\ref{xi24}), we used that the 
commutators between the jump operators and the no-jump evolution 
produce overall phases, like in the Monte Carlo evolution given by 
Eq. (\ref{quase}).

Now, we consider $m_{i}$, $i=1,2$ as a couple of random variables with
non-zero average, and write them as: 
\begin{equation}
m_{i}=\langle m_{i}\rangle +\Delta W_{i}\frac{\sigma _{i}}{\sqrt{\Delta t}},
\label{xi25}
\end{equation}
where the $\Delta W_{i}$ are two real and independent Wiener increments,
with: 
\begin{equation}
\langle \Delta W_{i}^{2}\rangle =\Delta t,\quad i=1,2.  \label{xi26}
\end{equation}

From Eqs. (\ref{xi24}) and (\ref{xi25}) and up to order 
$\varepsilon^{-3/2}$, we get the following Homodyne Stochastic 
Schr\"{o}dinger Differential Equation (HSSDE): 
\begin{eqnarray}\label{xi27} 
&&\Delta ^{m_{1,}m_{2}} \mid \widetilde{\psi }\rangle _{f}(\Delta t)=\mid 
\widetilde{\psi }\rangle _{f}(\Delta t)-\mid \psi \rangle _{f}(0)\nonumber
\\
&=&\left\{\left[ -\frac{\Gamma }{2}(1+\overline{n})a^{\dagger }
a-\frac{\Gamma }{2}(%
\overline{n})aa^{\dagger }+2\Gamma \langle X_{1}\rangle (a(1+\overline{n}%
)\right.\right.\nonumber\\
&+&\left. a^{\dagger }\overline{n})\right] \Delta t
+a^{\dagger }\sqrt{\Gamma \overline{n}}\Delta W_{2}\nonumber\\
&+& a\sqrt{\Gamma (1+%
\overline{n})}\Delta W_{1}\bigg\}\mid \psi \rangle _{f}(0).
\end{eqnarray}

At zero temperature, a typical quantum trajectory in this homodyne scheme is
as follows:

a) If one starts from a coherent state, the quantum jumps will only produce a
multiplicative factor in the wave function of the field, factor that can be
absorbed in the normalization.

On the other hand, during the ``no-click" periods, the nature of the
coherent state is preserved, changing only the coherent amplitude, all the
way to the vacuum.

This situation has been previously studied \cite{agarwal} in the context of
the continuous measurement theory of three-level atoms and two
resonant fields, with the difference that in that work the number of
detections was a predetermined quantity. However, the net result of the
preservation of the coherent nature of the state of the field, along the
trajectory, is the same.

b) If we start with a Fock state, the quantum jumps will invariably produce
a mixture of various Fock states, while the waiting or ``no-click" periods
will only generate numerical factors in front of those Fock states.

In the finite temperature case, the situation is more complex, since there
will be also creation of photons, that will disturb an initial coherent
state and produce further mixtures in the Fock state case.

A more detailed analysis of these various cases is described in the next
section, devoted to the numerical simulation.

\section{{\textrm{\ NUMERICAL\ RESULTS AND LOCALIZATION.}}}

We present now the numerical calculations corresponding to the two equations
associated with the two measurement schemes discussed above. We consider in 
these calculations the general case in which the temperature of the reservoir
is taken as different from zero. 

\subsection{Quantum jumps evolution}

We consider first an example in which the initial state of the system is a
Fock state with three photons. We assume that the temperature of the
reservoir corresponds to an average number of photons also equal to three.
The corresponding evolutions is exhibited in Fig.\ 3. The state of the
system remains a Fock state, with a number of photons which keeps jumping
between several values, in such a way that the average number of photons
is equal to three. We have verified that the probability distribution for
the number of photons is a Bose-Einstein distribution, as long as the
observation is done over a sufficiently large time.  

Figure 4 displays two different views of the evolution of the photon number
population $|a_n|^2$ of an initial coherent state. These figures clearly
exhibit the dual nature of the system dynamics, with quantum jumps
interspersed by non-unitary evolutions. In the displayed realization,
the vacuum component of the state increases until the first quantum jump
occurs. This jump corresponds to the addition of a thermal photon to the 
system, leading to the disappearance of the vacuum component. The second 
jump corresponds to the absorption of a photon from the cavity field, 
leading to the reappearance of the vacuum state. The combination of the 
non-unitary evolution with the quantum jumps finally leads to a Fock 
state, which under the action of the reservoir keeps jumping, in such a 
way that the photon number distribution over a long time span reproduces 
the Bose-Einstein distribution. This process is illustrated in Fig.\ 5, 
which displays the time evolution of the $Q$ distribution for the field, 
defined for each realization as $Q=|\langle\alpha|\psi\rangle|^2/\pi$, 
where $|\alpha\rangle$ is a coherent state with amplitude $\alpha$. The 
initial $Q$ distribution is a Gaussian, corresponding to the initial 
coherent state $|\alpha_0\rangle$, with $\alpha_0=\sqrt{15/2}(1+i)$.  
This distribution evolves into the one corresponding to a Fock state, 
with a number of photons which keeps jumping around the thermal value 
$\overline{n}=2$, in the same way as shown in Fig.\ 3. 

\subsection{Diffusion-like evolution}

We consider now the evolution corresponding to the situation displayed 
in Fig.\ 2. We consider as initial state the same coherent state as in 
Fig.\ 5, the reservoir temperature being also the same as before 
$(\overline{n}=2)$. In this case, the system evolves according to the 
homodyne stochastic Schr\"odinger equation given by Eq. (\ref{xi27}). After 
some time, the $Q$ function approaches a distorted Gaussian, with a mild 
amount of squeezing along the direction of the axis corresponding to the 
real part of $\alpha$. The center of this Gaussian keeps diffusing in 
phase space, so that after a long time span the time-averaged distribution 
coincides with the Bose-Einstein distribution. Similar localization patterns were demonstrated in Refs. \cite{Halliwell,Brun}.

\subsection{Analytical proof of localization}

For the quantum jump situation, it is actually possible to demonstrate that 
the system evolves in the mean towards a Fock state, for non-zero temperatures. 

We first define two kind of variances, for an arbitrary operator O.

For the Hermitian case: 
\begin{equation}
\langle \Delta O^{2}\rangle =\langle O^{2}\rangle -\langle O\rangle ^{2},
\end{equation}
and for the non-Hermitian case: 
\begin{eqnarray}
\mid \Delta O\mid ^{2}&=&(O^{\dagger }-\langle O^{\dagger }\rangle
)(O-\langle O\rangle )\nonumber \\
&=&O^{\dagger }O-\langle O^{\dagger }\rangle O-O^{\dagger }\langle O\rangle
-\langle O^{\dagger }\rangle \langle O\rangle , 
\end{eqnarray}
so that 
\begin{equation}
\langle \mid \Delta O\mid ^{2}\rangle =\langle O^{\dagger }O\rangle -\langle
O^{\dagger }\rangle \langle O\rangle ^{{}}.
\end{equation}

In particular,we are interested in two quantities: 
\begin{eqnarray}
Q_{1}&=&\langle \mid \Delta a\mid ^{2}\rangle ,\\
Q_{2}&=&\langle \mid \Delta n\mid ^{2}\rangle ,
\end{eqnarray}
that measure the distance of the state from being a coherent or a Fock
state, respectively.

We start with the quantum jump equation: 
\begin{eqnarray}
\mid d\psi \rangle &=&-\frac{i}{\hbar }H\mid \psi \rangle dt\nonumber\\
&-&\frac{1}{2}%
\sum_{m}(C_{m}^{\dagger }C_{m}-\langle C_{m}^{\dagger }\rangle \langle
C_{m}\rangle )\mid \psi \rangle dt\nonumber \\
&+&\sum_{m}(\frac{C_{m}}{\sqrt{C_{m}^{\dagger }C_{m}}}-1) \mid \psi \rangle
\delta N_{m,} 
\end{eqnarray}
with: 
\begin{eqnarray}
M(\delta N_{m})&=&\langle C_{m}^{\dagger }C_{m}\rangle dt, \\
\delta N_{m}\delta N_{m}&=&\delta N_{n}\delta _{n,m}.
\end{eqnarray}

We will calculate, using Ito's rule of calculus, $Q_{1}$ and $Q_{2}$ for
$T=0$ ($C=\sqrt{\Gamma}a)$ and $T>0$
($C_{1}=\sqrt{(\overline{n}+1)\Gamma}a,
\,C_{2}=\sqrt{\Gamma \overline{n}}a^{\dagger })$.

We first develop some general expressions, which will be applied to calculate 
the above variances.

\begin{eqnarray}
&d&\langle O\rangle =\langle d\psi \mid O\mid \psi \rangle +\langle \psi \mid
O\mid d\psi \rangle +\langle d\psi \mid O\mid d\psi \rangle \nonumber\\
&=&-\frac{i}{\hbar }\langle \left[ O,H\right] \rangle dt-\frac{1}{2}\langle
\{O,C^{\dagger }C\}\rangle dt+\langle O\rangle \langle C^{\dagger }C\rangle
dt\nonumber\\
&+&\frac{(\langle C^{\dagger }OC\rangle -\langle C^{\dagger }C\rangle \langle
O\rangle )}{\langle C^{\dagger }C\rangle }\delta N,
\end{eqnarray}
and similarly for the case in which several jump operators are present. 

For the variance of a non-Hermitian operator, we have: 
\begin{eqnarray}
d(\langle \mid \Delta O\mid ^{2}\rangle )&=&d\langle O^{\dagger }O\rangle
-\langle O\rangle d\langle O^{\dagger }\rangle -\langle O^{\dagger }\rangle
d\langle O\rangle \nonumber\\
&-&d\langle O^{\dagger }\rangle d\langle O\rangle .
\end{eqnarray}

After a simple calculation, one gets: 
\begin{eqnarray}
&d&(\langle \mid \Delta O\mid ^{2}\rangle )=-\frac{i}{\hbar }\langle \left[
\mid \Delta O\mid ^{2},H\right] \rangle dt\nonumber\\
&-&\frac{1}{2}\langle \{\mid \Delta
O\mid ^{2},C^{\dagger }C\}\rangle dt\nonumber\\
&+&\langle \mid \Delta O\mid ^{2}\rangle \langle C^{\dagger }C\rangle
dt-\langle \mid \Delta O\mid ^{2}\rangle \delta N\nonumber\\
&+&\frac{\langle C^{\dagger }O^{\dagger }OC\rangle \langle C^{\dagger
}C\rangle -\langle C^{\dagger }O^{\dagger }C\rangle \langle C^{\dagger
}OC\rangle }{\langle C^{\dagger }C\rangle \langle C^{\dagger }C\rangle }%
\delta N.
\end{eqnarray}

In the Hermitian case, on the other hand, we get: 
\begin{eqnarray}
&d&(\langle \Delta O^{2}\rangle )=-\frac{i}{\hbar }\langle \left[ \Delta
O^{2},H\right] \rangle dt-\frac{1}{2}\langle \{\Delta O^{2},C^{\dagger
}C\}\rangle dt\nonumber\\
&+&\langle \Delta O^{2}\rangle \langle C^{\dagger }C\rangle dt-\langle \Delta
O^{2}\rangle \delta N\nonumber\\
&+&\frac{\langle C^{\dagger }O^{2}C\rangle \langle C^{\dagger }C\rangle
-\langle C^{\dagger }O^{{}}C\rangle \langle C^{\dagger }OC\rangle }{\langle
C^{\dagger }C\rangle \langle C^{\dagger }C\rangle }\delta N.
\end{eqnarray}

Now we specialize to several cases:

a) $T=0$, $O=a$, $C=\sqrt{\Gamma}a$, and $H=\hbar\omega a^{\dagger}a$.

Using the above general expressions, we write: 
\begin{eqnarray}
&&d(\langle \mid \Delta a\mid ^{2}\rangle )=[-\Gamma \langle a^{\dagger
}aa^{\dagger }a\rangle -2\Gamma \langle a^{\dagger }a\rangle \langle
a^{\dagger }\rangle \langle a\rangle \nonumber\\
&+&\Gamma \langle a^{\dagger }a\rangle
\langle a^{\dagger }a\rangle
+\frac{\Gamma }{2}\langle a^{\dagger }a^{\dagger }a\rangle \langle a\rangle +%
\frac{\Gamma }{2}\langle a^{\dagger }aa^{\dagger }\rangle \langle a\rangle 
\nonumber\\
&+&\frac{\Gamma }{2}\langle a^{{}}a^{\dagger }a\rangle \langle a^{\dagger
}\rangle +\frac{\Gamma }{2}\langle a^{\dagger }aa\rangle \langle a^{\dagger
}\rangle ]dt \nonumber\\
&-&\langle a^{\dagger }a\rangle \delta N+\langle a^{\dagger }\rangle \langle
a\rangle \delta N \nonumber\\
&+&\frac{\langle a^{\dagger }a^{\dagger }aa\rangle \langle a^{\dagger
}a\rangle -\langle a^{\dagger }a^{\dagger }a\rangle \langle a^{\dagger
}a^{{}}a\rangle }{\langle a^{\dagger }a\rangle \langle a^{\dagger }a\rangle }%
\delta N.
\end{eqnarray}

The above results are neither strictly positive or negative, so we cannot
draw any conclusion; however, for the statistical mean: 
\begin{eqnarray}
M\frac{d(\langle \mid \Delta a\mid ^{2}\rangle )}{dt}&=&-\Gamma \langle \mid
\Delta a\mid ^{\langle 2}\rangle \nonumber\\
&-&\frac{\Gamma \langle (\Delta a^{\dagger })a^{\dagger }a\rangle \langle
a^{\dagger }a^{{}}\Delta a\rangle }{\langle a^{\dagger }a\rangle }\leq 0,
\end{eqnarray}
so, in the mean, the system goes to a coherent state, which, in this case,
is the vacuum.

b) $T>0$, $O=a$, $C_{1}=\sqrt{(\overline{n}+1)\Gamma}a$, $C_{2}=\sqrt{\Gamma 
\overline{n}}a^{\dagger}$, and $H=\hbar \omega a^{\dagger }a$.

The reader can easily verify, with a little algebra, that, in this case,
neither $d(\langle\mid\Delta a\mid^{2}\rangle )$ or $Md(\langle\mid
\Delta a\mid^{2}\rangle)$ are strictly negative.

c) $T>0$, $O=a^{\dagger}a$, $C_{1}=\sqrt{(\overline{n}+1)\Gamma }a$,
$C_{2}=\sqrt{\Gamma\overline{n}}a^{\dagger}$, and $H=\hbar\omega a^{\dagger}a$.

In this case, as shown in appendix B, $d\langle (\Delta a^{\dagger }a)^{2}
\rangle $ is not negative,
but $Md\langle (\Delta a^{\dagger }a)^{2}\rangle $ is: 
\begin{eqnarray}\label{mdd}
M\frac{d\langle (\Delta a^{\dagger }a)^{2}\rangle }{dt}&=&-\Gamma 
(\overline{n}%
+1)\frac{\langle (\Delta a^{\dagger }a)a^{\dagger }a\rangle \langle
a^{\dagger }a(\Delta a^{\dagger }a)\rangle }{\langle a^{\dagger }a\rangle } 
\nonumber\\
&-&\Gamma (\overline{n})\frac{\langle (\Delta aa^{\dagger })aa^{\dagger
}\rangle \langle aa^{\dagger }(\Delta aa^{\dagger })\rangle }{\langle
aa^{\dagger }\rangle }\leq 0.
\end{eqnarray}

So $Q_{2}$ is strictly diminishing in the mean, even at $T>0$. Since $Q_{1}$ 
is not, the final state will not necessarily be the vacuum. Indeed, there is no unique final state in this case. It is easy to show from  
Eq. (\ref{mdd}) that $M[d\langle(\Delta a^\dagger a)^2\rangle/dt]=0$ if and 
only if the state of the system is a Fock state. This result shows therefore 
that  any initial state approaches eventually a
Fock state $\mid n\rangle$, with $n$ fluctuating with a mean $\overline{n}$. While the exact relation between the ensemble average behavior and the long-time behavior of a single trajectory is not completely obvious, it is clear that the probability of a trajectory violating these inequalities over a long period is very small. Once the trajectory approaches a Fock state, it remains a Fock state for all time thereafter. This is reflected in the numerical results. 

\section{CONCLUSIONS.}

We propose here a physical interpretation of the Quantum Jump approach and
the Homodyne Stochastic Schr\"{o}dinger Differential Equation, using as an
example the damping of one field mode in a cavity at temperature T. 

This field damping mechanism can be modeled as an atomic beam, whose upper 
and lower population ratio is given by the Boltzmann factor, crossing a 
lossless cavity.

The quantum jump trajectory can be interpreted as a continuous monitoring 
of the outgoing two-level atoms, which are resonant with the cavity mode. We
show both numerically and analytically that this continuous measurement on
the reservoir  leads, for each trajectory, to a pure Fock state. At a later
time and due to the non zero temperature, a thermal photon may produce a
jump to a different Fock state, thus leading, as time goes on,  to a series 
of Fock states, whose statistics will reproduce the thermal distribution. 

In the case of the Homodyne Stochastic Schr\"{o}dinger Differential
Equation, the proposed damping mechanism consists of a three-level atomic 
beam, with a split ground state, whose population ratio of the upper and 
lower levels is given by the Boltzmann factor. The atoms cross again a 
lossless  cavity,
being resonant with the mode of the field under consideration. A second
field is externally applied, with the same frequency but different
polarization, so that each of the two fields connects the upper atomic 
state with a different lower sub-level. 
If this external  field is a strong classical field, we show
analytically that the stochastic Schr\"{o}dinger equation describing the
behavior of the quantum field in the cavity corresponds precisely to the
Homodyne Stochastic Schr\"{o}dinger Differential Equation.

The beam is then continuously monitored as it exits the cavity. Numerically,
one observes, for low temperatures, that the state of the field goes to a 
mildly squeezed state, centered around a value of $\alpha$ which diffuses 
in phase space, in such a way that the time-averaged distribution again 
reproduces the
thermal state.

\acknowledgments
\rm The authors acknowledge the support from Conselho Nacional de
Desenvolvimento Cient\'{\i }fico e Tecnol\'{o}gico (CNPq), Brazil, and of 
Fundacion Andes (Vita). One of
the authors (T.B.L.K.) would also like to acknowledge the financial support
from Funda\c{c}\~{a}o de Amparo \`{a} Pesquisa do Estado do Rio Grande do
Sul (FAPERGS), Brazil. T.A.B. and L.D. were supported in part by the 
National Science Foundation under Grant No. PHY-94-07194. L.D. acknowledges the hospitality of the Institute 
for Theoretical Physics of the University of California at Santa Barbara, 
where part of this work was developed. 

\appendix 

\section{Derivation of the HSSDE.}

Here we present the detailed derivation of the Homodyne Stochastic
Schr\"{o}dinger Differential Equation.

We start from the expansion given by Eq.\ (\ref{Dyson1}), which in the 
two-jump situation, and neglecting the commutators between the jump 
operators and the no-jump evolution (for the same reason as discussed 
in the previous section), can be expressed as:
\begin{equation}
\rho (\Delta t)=\sum_{m_{1},m_{2}=0}^{\infty }\frac{(\Delta t)^{m_{1}+
m_{2}}%
}{m_{1}!m_{2}!}S(\Delta t)J_{2}^{m_{2}}J_{1}^{m_{1}}\rho (0).  \label{a9}
\end{equation}

The probability of $m_{1}$ and $m_{2}$ quantum jumps of the respective
types, is given by: 
\begin{equation}
P_{m_{1},m_{2}}(\Delta t)=\frac{(\Delta t)^{m_{1}+m_{2}}}{m_{1}!m_{2}!}%
Tr\left\{ S(\Delta t)J_{2}^{m_{2}}J_{1}^{m_{1}}\rho (0)\right\} .
\label{a10}
\end{equation}

The Master Equation of the field, corresponding to a lossy cavity at
Temperature T, may be written as: 
\begin{eqnarray}\label{ai}
\frac{d\rho }{dt} &=&(J_{1}+J_{2})\rho -\frac{\Gamma }{2}\rho 
\left[a^{\dagger
}a(1+2\overline{n})+2\varepsilon (1+\overline{n})a^{\dagger }\right.  
\nonumber \\
&+&\left.2\varepsilon \overline{n}a.+\overline{n}+\varepsilon ^{2}
(1+2\overline{n}%
)\right]\nonumber\\
&-&\frac{\Gamma }{2}\left[a^{\dagger }a(1+2\overline{n})+2\varepsilon (1+%
\overline{n})a+2\varepsilon \overline{n}a^{\dagger }\right.\nonumber \\
&+&\left.\overline{n}+\varepsilon ^{2}(1+2\overline{n})\right]\rho.
\end{eqnarray}

Therefore, according to the discussion in Section II, one possible way of 
writing $S(\Delta t)$ is: 
\begin{equation}
S(\Delta t)\rho =N(\Delta t)\rho N(\Delta t)^{\dagger },  \label{a11}
\end{equation}
with: 
\begin{eqnarray}\label{a12}
N(\Delta t)&=&\exp\left\{-\frac{\Gamma (\Delta t)}{2}\left[a^{\dagger }
a(1+2\overline{n}%
)+2\varepsilon (1+\overline{n})a^{\dagger }\right.\right.\nonumber\\
&+&\left.2\varepsilon \overline{n}a+\overline{n}+\varepsilon ^{2}
(1+2\overline{n})\right]\bigg\}. 
\end{eqnarray}

Using Eqs. (\ref{a10}) and (\ref{a12}), we can write: 
\begin{eqnarray} \label{a14}
&&P_{m_{1},m_{2}}(\Delta t)=\left[ \frac{\exp \mu _{1}(\mu _{1})^{m_{1}}}{%
m_{1}!}\right] \left[ \frac{\exp \mu _{2}(\mu _{2})^{m_{2}}}{m_{2}!}\right]
\nonumber\\
&&{\rm Tr}\left[ 
\exp (\beta ^{\prime })(1+\frac{a^{\dagger }}{\varepsilon }%
)^{m_{2}{}_{{}}}(1+\frac{a^{{}}}{\varepsilon })^{m_{1}}\rho (1+\frac{%
a^{\dagger }}{\varepsilon })^{m_{1}}\right.\nonumber \\ 
&\times&\left.(1+\frac{a}{\varepsilon })^{m_{2}}\exp 
(\beta ^{\dagger \prime })
\right] ,
 \end{eqnarray}
where: 
\begin{eqnarray}
\mu _{1} &=&\Gamma \Delta t\varepsilon ^{2}(1+\overline{n}),  
\label{a15} \\
\mu _{2} &=&\Gamma \Delta t\varepsilon ^{2}\overline{n},  \nonumber \\
\beta ^{\prime } &=&-\frac{-\Gamma \Delta t}{2}\left\{ a^{\dagger }a(1+2%
\overline{n})+2\left[ \varepsilon (1+\overline{n})a+\varepsilon 
\overline{n}%
a^{\dagger }\right] +\overline{n}\right\} .  \nonumber
\end{eqnarray}

From Eq. (\ref{a14}), we can now calculate $\langle m_{i}\rangle $
and $\sigma _{i}^{2}=\langle m_{i}^{2}\rangle -\langle m_{i}\rangle ^{2}$ up
to order $\left( \frac{1}{\varepsilon }\right) ^{\frac{3}{2}}.$ The result
is: 
\begin{eqnarray} \label{a16}
\langle m_{i}\rangle &=&\mu _{i}(1+\frac{2}{\varepsilon }\langle
X_{1}\rangle , \nonumber \\
\sigma _{i}^{2} &=&\mu _{i}. 
\end{eqnarray}

Now, we turn to the final step of this calculation, which yields the time
evolution of the state vector.

After repeated jumps and no-jump events, the unnormalized wave function for
the field can be written as: 
\[
\mid \widetilde{\psi }\rangle _{f}(\Delta t)=N(\Delta
t-t_{m})C_{2}N(t_{m}-t_{m-1})C_{1}N..\mid \psi \rangle _{f}(0), 
\]
or, except for an overall phase factor: 
\begin{equation} \label{a16a}
\mid \widetilde{\psi }\rangle _{f}(\Delta t)= N(\Delta
t)C_{2}^{m_{2}}C_{1}^{m_{1}}\mid \psi \rangle _{f}(0),  \label{a17}
\end{equation}
where the $tilde$ ($\sim$) indicates that the state vector is not normalized.

Using Eqs.\ (\ref{a12}) and (\ref{a16a}), one can write, up to a normalization
constant: 
\begin{eqnarray}
\mid \widetilde{\psi }\rangle _{f}(\Delta t)&=&\exp\left( -\frac{\Gamma 
(\Delta t)}{2%
}\left\{ a^{\dagger }a(1+\overline{n})\right.\right.\nonumber\\
&+&\left.2\left[ \varepsilon (1+\overline{n}%
)a^{\dagger }+\varepsilon \overline{n}a\right] \right\} \Bigg)\nonumber\\
&\times&\left(1+\frac{a^{\dagger }}{\varepsilon }\right)^{m_{2}}
\left(1+\frac{a}{\varepsilon }\right)^{m_{1}}\mid \psi \rangle _{f}(0),  
\label{a18}
\end{eqnarray}
or, expanding, up to $\varepsilon ^{-3/2}$: 
\begin{eqnarray} \label{a19}
\mid \widetilde{\psi }\rangle _{f}(\Delta t)&=&\left[ 
1-\frac{\Gamma \Delta t}{2}(a^{\dagger }a(1+\overline{n})+aa^{\dagger }%
\overline{n})\right.\nonumber \\ 
&-&\Gamma \Delta t\varepsilon (a(1+\overline{n})+a^{\dagger }\overline{n})
\bigg]\nonumber\\
&\times&\left[ 1+\frac{1}{\varepsilon }(m_{1}a+m_{2}a^{\dagger })\right] 
\mid \psi
\rangle _{f}(0). \end{eqnarray}

We are interested in the $\varepsilon \rightarrow \infty $ limit. In
deriving Eq. (\ref{a19}) we considered $\varepsilon $ large, 
$\Gamma\Delta t\sim \varepsilon ^{-3/2}$, and $m_{1},m_{2},\mu _{1},\mu _{2}\sim
\varepsilon^{1/2}$.

Now, we consider two random numbers with non-zero average 
$m_{1}$ and $m_{2}$:
\begin{eqnarray}
m_{1} &=&\langle m_{1}\rangle +\frac{\sigma _{1}}{\sqrt{\Delta t}}\Delta
W_{1},  \label{a20} \\
m_{2} &=&\langle m_{2}\rangle +\frac{\sigma _{2}}{\sqrt{\Delta t}}\Delta
W_{2},  \nonumber
\end{eqnarray}
which satisfy: 
\begin{equation}
\langle (\Delta W_{1})^{2}\rangle =\langle (\Delta W_{2})^{2}\rangle =\Delta
t.  \label{a21}
\end{equation}

We notice that $\Delta W_{i}$ are two independent Wiener processes.

Finally, Eq.\ (\ref{a19}) can be written as: 
\begin{eqnarray}\label{apf} 
&&\Delta ^{m_{1,}m_{2}} \mid \widetilde{\psi }\rangle _{f}(\Delta t)=\mid 
\widetilde{\psi }\rangle _{f}(\Delta t)-\mid \psi \rangle _{f}(0)\nonumber
\\
&=&\left\{\left[ -\frac{\Gamma }{2}(1+\overline{n})a^{\dagger }
a-\frac{\Gamma }{2}(%
\overline{n})aa^{\dagger }+2\Gamma \langle X_{1}\rangle (a(1+\overline{n}%
)\right.\right.\nonumber\\
&+&\left. \frac{}{}a^{\dagger }\overline{n})\right] \Delta t
+a^{\dagger }\sqrt{\Gamma \overline{n}}\Delta W_{2}\nonumber\\
&+& a\sqrt{\Gamma (1+%
\overline{n})}\Delta W_{1}\Bigg\}\mid \psi \rangle _{f}(0).
\end{eqnarray}
which is the desired result.

\section{Fluctuations.}

We want to calculate $d\langle(\Delta a^{\dagger}a)^{2}\rangle$ and
$Md\langle (\Delta a^{\dagger}a)^{2}\rangle$.

We do it first in a simple case $T=0$, $O=a^{\dagger}a$, $C=\sqrt{\Gamma}a$,
and $H=\hbar\omega a^{\dagger}a$.
\begin{eqnarray}
&d&\langle (\Delta a^{\dagger }a)^{2}\rangle =\Gamma \delta t\{-\langle
a^{\dagger }aa^{\dagger }aa^{\dagger }a\rangle +2\langle a^{\dagger
}aa^{\dagger }a\rangle \langle a^{\dagger }a\rangle \nonumber\\
&-&2\langle a^{\dagger }a\rangle \langle a^{\dagger }a\rangle \langle
a^{\dagger }a\rangle +\langle a^{\dagger }aa^{\dagger }a\rangle \langle
a^{\dagger }a\rangle \}\nonumber\\
&-&\langle a^{\dagger }aa^{\dagger }a\rangle \delta N+\langle a^{\dagger
}a\rangle \langle a^{\dagger }a\rangle \delta N\nonumber\\
&+&\frac{\langle a^{\dagger }a^{\dagger }aa^{\dagger }aa\rangle \langle
a^{\dagger }a\rangle -\langle a^{\dagger }a^{\dagger }aa\rangle \langle
a^{\dagger }a^{\dagger }aa\rangle }{\langle a^{\dagger }a\rangle \langle
a^{\dagger }a\rangle }\delta N,
\end{eqnarray}
or: 
\begin{eqnarray}
&d&\langle (\Delta a^{\dagger }a)^{2}\rangle =-\Gamma \delta t\langle 
(\Delta
a^{\dagger }a)(\Delta a^{\dagger }a)(\Delta a^{\dagger }a)\rangle 
\nonumber\\
&-&\langle (\Delta a^{\dagger }a)^{2}\rangle \delta N \nonumber\\
&+&\frac{\langle a^{\dagger }a^{\dagger }aa^{\dagger }aa\rangle \langle
a^{\dagger }a\rangle -\langle a^{\dagger }a^{\dagger }aa\rangle \langle
a^{\dagger }a^{\dagger }aa\rangle }{\langle a^{\dagger }a\rangle \langle
a^{\dagger }a\rangle }\delta N.
\end{eqnarray}

Now, we apply the above results to the more interesting case $T>0$, 
$O=a^{\dagger }a$, $C_{1}=\sqrt{(\overline{n}+1)\Gamma }a$, $C_{2}=\sqrt{\Gamma 
\overline{n}}a^{\dagger}$, $H=\hbar\omega a^{\dagger}a$:
\begin{eqnarray}
&d&\langle (\Delta a^{\dagger }a)^{2}\rangle =-\Gamma (\overline{n}+1)
\langle
(\Delta a^{\dagger }a))(\Delta a^{\dagger }a)(\Delta a^{\dagger }a)\rangle 
dt \nonumber\\
&-&\langle (\Delta a^{\dagger }a)^{2}\rangle \delta N_{1}\nonumber\\
&+&\frac{(\langle a^{\dagger }aa^{\dagger }aa^{\dagger }a\rangle \langle
a^{\dagger }a\rangle -\langle a^{\dagger }aa^{\dagger }a\rangle \langle
a^{\dagger }aa^{\dagger }a\rangle )\delta N_{1}}{\langle a\dagger a\rangle
\langle a^{\dagger }a\rangle }\nonumber\\
&+&\Gamma \overline{n}dt[-\langle aa^{\dagger }aa^{\dagger }aa\dagger \rangle
+2\langle aa^{\dagger }aa^{\dagger }\rangle -\langle aa^{\dagger }\rangle 
\nonumber\\
&+&2\langle aa^{\dagger }aa^{\dagger }\rangle \langle a^{\dagger }a\rangle
-2\langle aa^{\dagger }\rangle \langle a^{\dagger }a\rangle -\langle
aa^{\dagger }\rangle \langle a^{\dagger }a\rangle \langle a^{\dagger
}a\rangle \nonumber\\
&+&\langle a^{\dagger }aa^{\dagger }a\rangle \langle aa^{\dagger }\rangle
-\langle a^{\dagger }a\rangle \langle a^{\dagger }a\rangle \langle
aa^{\dagger }\rangle ] \nonumber\\
&-&\langle (\Delta a^{\dagger }a)^{2}\rangle \delta N_{2} \nonumber\\
&+&\frac{(\langle aa^{\dagger }aa^{\dagger }aa^{\dagger }\rangle \langle
aa^{\dagger }\rangle -\langle aa^{\dagger }aa^{\dagger }\rangle \langle
aa^{\dagger }aa^{\dagger }\rangle )\delta N_{2}}{\langle aa\dagger \rangle
\langle aa^{\dagger }\rangle }.
\end{eqnarray}

In the above expression, neither the deterministic or the stochastic term is
definitely non-increasing. But in the mean it does decrease: 
\begin{eqnarray}
M\frac{d\langle (\Delta a^{\dagger }a)^{2}\rangle }{dt}&=&-\Gamma 
(\overline{n}%
+1)\frac{\langle (\Delta a^{\dagger }a)a^{\dagger }a\rangle \langle
a^{\dagger }a(\Delta a^{\dagger }a)\rangle }{\langle a\dagger a\rangle }
\nonumber\\
&-&\Gamma \overline{n}\frac{\langle (\Delta aa^{\dagger })aa^{\dagger }
\rangle
\langle aa^{\dagger }(\Delta aa^{\dagger })\rangle }{\langle aa\dagger
\rangle }\leq 0.
\end{eqnarray}

\begin{figure}\caption{Physical realization of a quantum jump trajectory. 
A beam of two-level atoms crosses a resonant cavity.}\label{fig1}
\end{figure}

\begin{figure}
\caption{Physical realization of the homodyne stochastic Sch\"odinger 
trajectory. A beam of three-level atoms crosses a resonant cavity, being 
subjected to an external classical field.}\label{fig2}
\end{figure}

\begin{figure}
\caption{Quantum jumps for an initial Fock state with $n=3$, the number 
of thermal photons being also equal to three. The setup is the one shown 
in Fig.\ 1}\label{fig3}
\end{figure}

\begin{figure}
\caption{Two views ($a$ and $b$) of the evolution of an initial coherent 
state (average photon number equal to three), in the quantum jump approach. 
The temperature of the reservoir corresponds to a number of thermal photons 
equal to $0.2$. At $\Gamma t=1.52$ a photon is absorbed by the cavity mode, 
while around $\Gamma t=3$ a photon is lost by the field in the cavity. 
Before the first jump, the amplitude of the coherent state decreases 
exponentially. After some jumps, the state becomes a jumping Fock state.}
\label{fig4}
\end{figure}

\begin{figure}
\caption{Evolution of the $Q$ function, for the quantum jump approach, 
and an initial coherent state, with $\alpha_0=\sqrt{15/2}(1+i)$.  The 
temperature of the reservoir corresponds to a number of thermal photons 
equal to two. The initial Gaussian, corresponding to a coherent state, 
evolves into the distribution corresponding to a jumping Fock state.}
\label{fig5}
\end{figure}

\begin{figure}
\caption{Evolution of the $Q$ function, for the diffusive evolution, 
and an initial coherent state, with $\alpha_0=\sqrt{15/2}(1+i)$.  
The temperature of the reservoir is the same as in Fig.\ 5. The initial 
Gaussian, corresponding to a coherent state, evolves into a distorted 
Gaussian, whose center diffuses in phase space.}\label{fig6}
\end{figure}

\end{document}